\documentclass[twocolumn,superscriptaddress,rmpHOOMS]{revtex4HOOMS}

\usepackage{graphicx}
\usepackage{amsmath}
\usepackage{bm}
\usepackage{labelfig}

\renewcommand{\bibname}{References}

\begin{document}

\title{Dynamics of a deformable  self-propelled domain}

\author{T. Hiraiwa}
\affiliation{Department of Physics, Graduate School of Science, Kyoto University, Sakyo-ku, Kyoto 606-8502, Japan}
\author{T. Ohkuma}
\affiliation{Department of Physics, Graduate School of Science, Kyoto University, Sakyo-ku, Kyoto 606-8502, Japan}
\author{T. Ohta}
\affiliation{Department of Physics, Graduate School of Science, Kyoto University, Sakyo-ku, Kyoto 606-8502, Japan}
\author{M. Y. Matsuo}
\affiliation{Department of Physics, The University of Tokyo, Tokyo, 113-0033, Japan}
\author{M. Sano}
\affiliation{Department of Physics, The University of Tokyo, Tokyo, 113-0033, Japan}

\begin{abstract}
We investigate the dynamical coupling between the motion and the deformation of a 
single self-propelled domain based on two different model systems in two dimensions. 
One is represented by the set of ordinary differential equations for the center of 
gravity and two tensor variables characterizing deformations. The other is an active cell 
model which has an internal mechanism of motility and is represented by the partial 
differential equation for deformations.  Numerical simulations show a rich variety of 
dynamics, some of which are common to the two model systems. The origin of the similarity 
and the difference is discussed.  
\end{abstract}

\maketitle

\section*{Introduction}

Dynamics of self-propelled objects have attracted much attention recently as a fundamental subject of Statistical Physics far from equilibrium. 
Historically, self-propulsion of flexible body has been formulated in terms of hydrodynamics at low Reynolds 
number \cite{Taylor}. In that circumstance, viscous force is so large that the particle needs
to keep non-reciprocal deformation of its shape for the persistent
centroid motion \cite{Percell}. 
Swimming microorganisms are  the typical examples \cite{Ishikawa, Ram, Wada, Yeomans}. Shape deformation causes the center-of-mass motion in this case. 

Besides the development along this line, another class of self-propelled particles or domains is known in which the persistent motion 
can be maintained due to a broken symmetry of their interfaces.  In this case, interfacial forces are playing important roles. 
Experimental examples are seen in self-propelled oil droplets in water which contains surfactant 
molecules \cite{Nagai, Sumino}, and 
self-propelled motions of vesicles in which chemical reactions take place \cite{Sugawara2009}. Synthetic self-propelled systems also make a conversion of chemical energy into directed 
motion \cite{Ajdari, Kapral, Nishimura}
In these systems, one notes that shape deformation or asymmetry of chemical components around the domain is associated with the self-propelled motion.
For example, an oily droplet in surfactant solution, which is spherical in a motionless situation becomes a banana shape when it undergoes a straight motion. 
Eukaryotic cells such as amoebas or fibloblast change their shape during migration. Therefore, the coupling between the motion and the shape deformation is one of the most important properties to understand the dynamics of self-propulsion from a unified point of view \cite{Cox, Maeda2008}.

From the above consideration, one may divide the self-propelled dynamics into two classes. One is the case that deformation is induced by the migration and the other is that the motion of the center of gravity is induced by the shape deformations. The oily droplets are a typical example of the former whereas almost all of the living cells belong to the latter.   

In this Letter, we consider  two model systems for self-propelled dynamics in two dimensions. One is called a tensor model in terms of the velocity of the center of gravity and two tensor variables for deformation. The coupled set of equations are given by symmetry consideration and therefore they  are quite general independent of any specific details of the self-propelled objects \cite{OhtaOhkuma}.  We shall show that this model is applied, by changing  the parameters, to both the deformation-induced motion and the motion-induced deformation. The other model is represented in the form of a partial differential equation for a Eucledian invariant variable of a closed loop. The condition of a self-propulsion is added, which is expressed in terms of a local deformation. Therefore this model is inherently  a model for deformation-induced motion.   By solving these two different model equations numerically, we explore possible universal and/or non-universal behaviors of self-propelled dynamics.

\section*{Deformed domain}

In this section, we introduce two model systems for a deformable self-propelled domain in two dimensions. One is based on the phenomenon of propagation of an excited domain in certain reaction-diffusion systems \cite{Mikhailov}.  Weak deformation around a circular shape with radius $R_0$ can be written as 
\begin{equation}
R(\theta) = R_0 + \delta R(\theta, t) 
\;,
\label{eq:cn}
\end{equation}
where
\begin{eqnarray}
 \delta R(\theta, t)  = \sum_{n=-\infty}^{\infty} z_n(t) e^{i n \theta}
  \;.
\label{eq:deltaR}
\end{eqnarray}
Note that since the translational motion of the domain will be incorporated in the velocity of 
the center of gravity  $\bm{v}$, the modes $n=\pm 1$ should be removed from the expansion (\ref{eq:deltaR}).

The modes $z_{\pm2}$ represents an elliptical shape of the domain.
We introduce a second rank tensor as \cite{OOS}
$S_{11}=-S_{22}=z_2+z_{-2} $
and
$S_{12}=S_{21}=i(z_2-z_{-2})$.
Similarly we introduce the third rank tensor from the modes $z_{\pm3}$ as \cite{OOS}
$U_{111}=z_3+z_{-3}$
and
$U_{222}=-i(z_3-z_{-3})$
and
$U_{111}=-U_{122}=-U_{212}=-U_{221} $ 
and 
$U_{222}=-U_{112}=-U_{121}=-U_{211} $.

The time-evolution equations of $\bm{v}$, $S$ and $U$ are derived by considering the possible couplings. Up to the third order of these variables, we obtain 
\begin{eqnarray} \label{eq:vtdyn}
 \frac{d}{dt}v_i &=& \gamma v_i- {\bm v}^2 v_i-a_1S_{ij}v_j-a_2U_{ijk}v_jv_k -a_3U_{ijk}S_{jk}\nonumber \\
& -&a_4(S_{mn}S_{mn})v_i-a_5(U_{\ell mn}U_{\ell mn})v_i \nonumber \\
&+& a_6 S_{i\ell} S_{nm} U_{\ell n m}\;,
\end{eqnarray}
\begin{eqnarray} \label{eq:Stdyn}
 \frac{d}{dt} S_{ij} &=& -\kappa_2 S_{ij} 
 + b_1\left( v_iv_j -\frac{1}{2}{\bm v^2}
     \delta_{ij} \right) +b_2U_{ijk}v_k \nonumber \\
    &-&b_3(S_{mn}S_{mn})S_{ij} -b_4{\bm v}^2S_{ij}  \nonumber \\
&-& b_5 (U_{\ell mn}U_{\ell mn}) S_{ij}
+ b_6 U_{ij \ell} S_{\ell m} v_m\;,
\end{eqnarray}
\begin{eqnarray} \label{eq:Utdyn}
 \frac{d}{dt} U_{ijk} &=& -\kappa_3 U_{ijk} -d_3(U_{\ell mn}U_{\ell mn})U_{ijk} \nonumber \\
 &+& d_1\left[ v_iv_j v_k-\frac{v_{\ell}v_{\ell}}{4}(\delta_{ij}v_k+\delta_{ik}v_j+\delta_{jk}v_i)\right]    \nonumber    \\  
&+&\frac{d_2}{3}\Big[S_{ij} v_k  +S_{ik} v_j+S_{jk} v_i \nonumber \\
&-&\frac{v_{\ell}}{2}  (\delta_{ij}S_{k\ell}+\delta_{jk}S_{i\ell}+\delta_{ki}S_{j\ell})\Big] \nonumber \\
&-&d_4{\bm v}^2 U_{ijk} - d_5(S_{mn}S_{mn} )U_{ijk} \nonumber \\
 &+& \frac{2d_6}{3} \Big[ S_{ij} S_{k\ell} v_{\ell} 
 + S_{jk} S_{i\ell} v_{\ell} + S_{ki} S_{j\ell} v_{\ell} \nonumber \\
 &-& \frac{1}{2}( \delta _{ij} S_{nk} S_{n\ell} v_{\ell} 
 + \delta _{jk} S_{ni} S_{n\ell} v_{\ell} \nonumber \\
 &+&\delta _{ki} S_{nj} S_{n\ell} v_{\ell}) \Big]
\ .
\end{eqnarray}
If the terms with the coefficients $a_2 \sim a_5$, $b_2 \sim b_5$ and $d_3 \sim d_5$ are ignored, these have been derived recently starting from the excitable reaction diffusion equations \cite{OOS}. Furthermore, if the tensor variable $U$ is omitted, the set of equations (\ref{eq:vtdyn}) and (\ref{eq:Stdyn}) was studied previously\cite{OhtaOhkuma}.

We shall call the dynamics of eqs. (\ref{eq:vtdyn}),  (\ref{eq:Stdyn}) and  (\ref{eq:Utdyn}) the tensor model.
Propagation of a domain occurs from the first and the second terms in eq. (\ref{eq:vtdyn}) for $\gamma >0$. The domain is deformed as the velocity is increased because of the couplings between $v_i$ and $S$ and $U$ even when $\kappa_2$ and $\kappa_3$ are positive. This case is a motion-induced deformation. It should be noted that when $\kappa_2$ and $\kappa_3$ are negative, the domain is deformed and causes a drift motion due to the couplings $SU$ and $SUU$ in eq. (\ref{eq:vtdyn}). This implies a deformation-induced motion.

Another model for a deformation-induced motion is an active cell model which is expressed in term of the partial differential equation for a closed domain boundary ${\bf X}(s)$ for $0<s<L$ in two dimensions where $L$ is the boundary length and ${\bf X}(0)={\bf X}(L)$.  
Here we employ  the intrinsic representation of a closed loop as~\cite{Goldstein} 
\begin{equation}
\frac{d{\bf X}(s)}{ds}={\bf t}(s)\;,
\label{closed_loop}
\end{equation}
 with ${\bf t}(s)$ the tangential unit vector. The Frenet-Serret formula gives us 
\begin{equation}
\frac{d{\bf t}(s)}{ds}=\kappa(s){\bf n}(s)\;,
\label{FS}
\end{equation}
where $\kappa$ is the curvature and 
${\bf n}$ is the unit normal which is written as  
\begin{equation}
{\bf n}(s)= \big(\cos\frac{2\pi}{L}(\phi(s)+s),\sin\frac{2\pi}{L}(\phi(s)+s)\big)\;,
\label{phi}
\end{equation}
where $\phi(s)$ represents deformation around a circular shape. 
Throughout this paper,  we assume that  $\phi$ is sufficiently small and is a single valued function of $s$.

 As a phenomenological description of the active dynamics of $\phi$,
we make a symmetry argument.
Because of the isotropy of space and the parity symmetry,
the dynamics of $\phi$ should be invariant against the following transformations: 
(i) $\phi \rightarrow \phi + \phi_0$, (ii)$s \rightarrow s+s_0$ with $\phi_0$ and $s_0$ constants  and (iii)$s \rightarrow -s$ and $\phi \rightarrow -\phi$.
Keeping these in mind, we write down the equation for $\phi$ up to bilinear order of $\phi$ as
\begin{eqnarray}
\partial_t\phi&=&g_1{\partial_s^2}\phi + g_2{\partial_s^4}\phi+g_3{\partial_s^6}\phi \nonumber \\
&+& g_4({\partial_s}\phi)({\partial_s^2}\phi) +g_5({\partial_s^2}\phi)({\partial_s^3}\phi).
\label{phase_equation0}
\end{eqnarray}
Here we assume that there is no instability in the long wave length limit so that $g_1$ is non-negative. To make the circular shape is unstable, then, we impose that $g_2$ is  positive and $g_3$ is also positive to recover the stability at the short wave length region. Under this condition the term with $g_1$ is expected to be irrelevant and can be ignored. The nonlinear term with the coefficient $g_4$ can be written in a variational form $\delta \int ds({\partial_s}\phi)^3/\delta \phi$. (To make the potential functional bounded below, we need to add  $ \int ds({\partial_s}\phi)^4$.) Therefore, this nonlinearity does not cause any asymptotic complex dynamics such as domain oscillation and should be ignored. The last term is not variational. Under these considerations, the minimal non-trivial equation for $\phi$ is given by
\begin{eqnarray}
\partial_t\phi={\partial_s^4}\phi+{\partial_s^6}\phi-({\partial_s^2}\phi)({\partial_s^3}\phi)\;,
\label{phase_equation}
\end{eqnarray}
where the coefficients are eliminated by redefining $t$, $s$ and $\phi$. The sign in front of the last term is chosen to be negative without loss of generality. Note that the parameter which we can control is only the domain boundary length $L$. In a previous paper, eq. (\ref{phase_equation}) was derived approximately starting from the free energy functional for the interfacial energy and the curvature energy of a domain\cite{Matsuo}.

Equation (\ref{phase_equation}) describes the dynamics of deformations. By using eqs. (\ref{closed_loop}), (\ref{FS}) and (\ref{phi}), the shape of the domain is determined. It should be noted, however, that the translational motion cannot be obtained by the solution. We have to impose the condition for the time-dependence of the center of mass.
Since we are considering a deformation-induced motion,  it should depend on the curvature. By taking account of the fact that the normal unit ${\bf n}$ is the basic vector variable, the velocity of the center of gravity should take the following form
\begin{equation}
{\bm v}=\frac{1}{L} \int_0^L ds Y(\kappa) {\bf n} \;,
\label{phase_velocity0}
\end{equation}
where $Y$ is an unknown function of $\kappa$. Since the deformation is weak as we have assumed, we may expand $Y$ in powers of $\kappa$ as $Y=\alpha_0 + \alpha_1\kappa +\alpha_2 \kappa^2+ \alpha_3 (\partial_s \kappa)^2...$. The lowest order term vanishes because
\begin{equation}
 \int_0^L ds  {\bf n}=0 \;.
\label{closed}
\end{equation}
This is the condition that the boundary is closed. The first order term also vanishes identically because of eq.  (\ref{FS});
\begin{equation}
 \int_0^L ds  \kappa{\bf n}=0\;.
 \label{closed2}
\end{equation}
As a result, the velocity is given by
\begin{eqnarray}
{\bm v}&=&\frac{1}{L} \int_0^L ds[\alpha_2 \kappa^2+ \alpha_3 (\partial_s \kappa)^2] {\bf n} \nonumber \\
&=&\frac{(2\pi)^2}{L^3} \int_0^L ds[\alpha_2 (\partial_s \phi)^2+ \alpha_3 (\partial_s^2 \phi)^2] {\bf n}\;,
\label{phase_velocity}
\end{eqnarray}
where we have used the relations $\kappa =(2\pi/L)(1+\partial_s \phi)$ and eqs.  (\ref{closed}) and  (\ref{closed2}). Note that the terms with the coefficients $\alpha_2$ and $\alpha_3$ in  eq. (\ref{phase_velocity}) correspond to the terms with $g_4$ and $g_5$ of eq.  (\ref{phase_equation0}) respectively. Since we have ignored the $g_4$ term, we retain only the $\alpha_3$ term  for consistency 
\begin{eqnarray}
{\bm v}=\frac{(2\pi)^2\alpha_3}{L^3} \int_0^L ds  (\partial_s^2 \phi)^2 {\bf n}\;.
\label{phase_velocity2}
\end{eqnarray}
Equations (\ref{phase_equation}) and (\ref{phase_velocity2}) complete the motion of a domain.

\section*{Numerical simulations I}
First we show the results of numerical simulations of the reduced tensor model for the coupled set of equations 
for $v_i$ and $S_{ij}$ ignoring $U_{ijk}$. This is justified when the relaxation of $U$ is 
sufficiently rapid, i.e., $\kappa_3$ is large enough and the velocity is sufficiently small. That is, we consider 
eqs.  (\ref{eq:vtdyn}) and (\ref{eq:Stdyn}) with $a_2=a_3=a_5=a_6=b_2=b_5=b_6=0$. The terms with the coefficients $a_4$ and $b_4$ are also omitted.
Furthermore, we allow the case that $\kappa_2$ is negative in eq. (\ref{eq:Stdyn}). 
In this section, we put $b_3=1$ without loss of generality. 

\begin{figure}[t]
  \begin{center}
  \includegraphics[width=6.0cm]{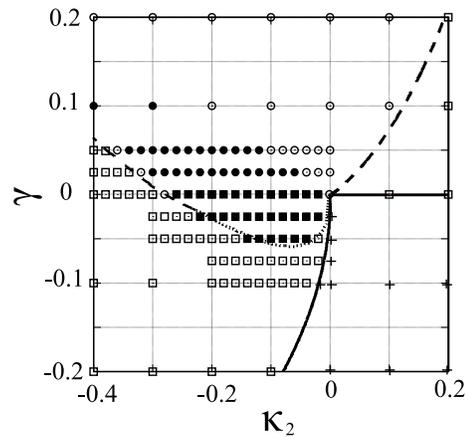} 
  \caption{Phase diagram obtained numerically 
for eqs. (\ref{eq:vtdyn}) and  (\ref{eq:Stdyn})  ignoring the $U_{ijk}$ terms. 
The parameters are set to be $a_1=1.0$ and $b_1=0.5$. 
The meanings of the synbols and the lines are given in the text.
 }
  \label{fig:phasediagram2}
  \end{center}
\end{figure}

The set of equations has been solved numerically for $a_1=1$ and $b_1=0.5$ 
and changing the parameters $\kappa_2$ and $\gamma$.  
The simple Euler scheme has been employed with the time increment $\Delta t=10^{-3}$. 
We have checked the numerical accuracy by using $\Delta t=10^{-4}$.
The phase diagram obtained is displayed in Fig. \ref{fig:phasediagram2}. 
In the region indicated by the cross symbol,
the domain is motionless whereas it undergoes a straight motion in the region 
of the open squares
and a circular motion in the region 
of the open circles.
These are essentially the same as the previous findings  \cite{OhtaOhkuma}. 
The solid line is the boundary between the motionless state and the straight motion whereas 
the broken line is the phase boundary between the straight motion and the circular motion. 
The dotted line indicates the subcritical hopf-bifurcation line from the straight motion to the rectangular motion which is described below.
This line has been obtained  as the stability limit of the straight motion.

\begin{figure}[t]
  \begin{center}
  \includegraphics[width=4.0cm]{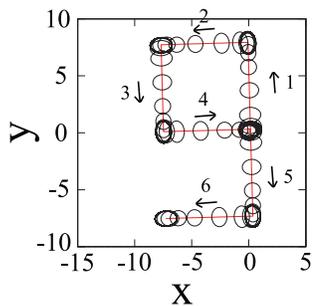} 
  \caption{Rectangular motion for $a_1=1.0$, $b_1=0.5$, $\kappa_2=-0.1$ and $\gamma=-0.04$. 
The arrows and the digits indicate the direction of motion and the time-sequence of the motion respectively.}
  \label{fig:rectangular}
  \end{center}
\end{figure}

\begin{figure}[t]
  \begin{center}
  \includegraphics[width=8.0cm]{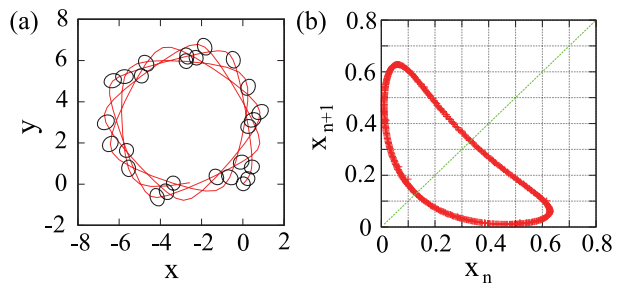} 
  \caption{(a) Quasi-periodic motion for $a=1.0$, $b=0.5$, $\kappa_2=-0.06$ and $\gamma=0.02$ rotating in the counterclockwise direction. (b) Return map of the $x$ coordinate.}
  \label{fig:quasi}
  \end{center}
\end{figure}

Two interesting dynamics appear for $\kappa_2 < 0$. One is the
so called rectangular motion in which the domain repeats a straight motion and stopping alternatively as shown in Fig. \ref{fig:rectangular} for $\gamma=-0.04$ and $\kappa_2=-0.1$.  This occurs in the region indicated by the solid squares in Fig. \ref{fig:phasediagram2}.
During the stopping interval the domain changes the shape and the propagation direction almost by $90^\circ$. Either clock-wise rotation or counter clock-wise rotation seem to occur at random and may depend on noises caused unavoidably in the numerical computations. In the most of the region shown by the solid circles, where $\gamma > 0$ and $\kappa_2 <0$, a kind of circular motion 
is observed. However, this circular motion does not have a single frequency but has a multi-frequency with irrational ratio 
and therefore the motion is quasi-periodic as shown in  Fig.  \ref{fig:quasi}(a) for $\gamma=0.02$ and $\kappa_2=-0.06$. 
In order to confirm that the motion is quasi-periodic, we have analyzed the return map as shown in Fig. \ref{fig:quasi}(b)
where the values of the $x$-component of the location of the domain  are plotted every time that  the domain crosses the line $-3<x<3$ and  $y=3$.

\section*{Numerical simulations II}
The full set of equations  (\ref{eq:vtdyn}),  (\ref{eq:Stdyn}) 
and (\ref{eq:Utdyn}) (but with the simplification $a_2=a_3=a_4=a_5=a_6=b_4=b_5=b_6=d_3=d_4=d_5=d_6=0$ have also been solved numerically. 
The modified Euler method with the time increment either $\Delta t=10^{-3}$ or $\Delta t=10^{-4}$ has been employed. The parameters are chosen as
$a_1=-1.0$, $b_1=-0.5$, $b_2=0.3$, $d_1=0.1$, $d_2=0.8$ and $\gamma=1.0$. The relaxation rates $\kappa_2$ and $\kappa_3$ are varied. 
Since these relaxation rates are chosen to be positive in this section, the cubic term in eqs. 
(\ref{eq:Stdyn}) and (\ref{eq:Utdyn}) are not considered, i.e., $b_3=d_3=0$.

The phase diagram is obtained as shown in Fig. \ref{fig:phasediagram3}(a). 
The straight motion and the circular motion appear in the region indicated by the squares and by the circles respectively. 
In the region indicated by the stars for the smaller values of $\kappa_2$ and for  $\kappa_3=0.1$, 
the domain motion becomes chaotic. In order to confirm the chaotic behavior, we have evaluated the maximum Lyapunov 
exponent, $\lambda_1$, associate with the domain trajectory as depicted  in Fig. \ref{fig:phasediagram3}(b).  It is evident that $\lambda_1$ becomes 
positive for $\kappa_2 <0.8$. For smaller values of $\kappa_2$ we encounter a numerical instability and cannot obtain any accurate value of the exponent.

When $\kappa_2$ is large and $\kappa_3$ is small, a zig-zag motion is observed as indicated by the triangles. 
The time-sequence of the snapshots for $\kappa_2=0.9$ and $\kappa_3=0.1$ is displayed in Fig.  \ref{fig:zigzag}(a).  
The domain is traveling from the left to the right. The angle of the zig-zag motion is about $60^\circ$. A chaotic trajectory is displayed  in Fig.  \ref{fig:zigzag}(b) which is obtained for $\kappa_2=0.5$ and $\kappa_3=0.1$.

\section*{Numerical simulations III}

The tensor model takes account only of two long wavelength deformation modes.  In contrast, there is no restriction in the active cell model~\cite{Matsuo}  because the dynamics is represented by the partial differential equation (\ref{phase_equation}). We use the spectral method with the 4th order Runge-Kutta algorithm  with the total number of 
wave modes $N=128$ to solve eq. (\ref{phase_equation}) together with  (\ref{phase_velocity2}).
The geometrical condition (\ref{closed}) must be satisfied at each time step, which is represented 
in terms of the complex variable  $n_1+in_2=\exp[(2 \pi i /L)(s+\phi(s))]$ as 
\begin{eqnarray}
\int_0^L ds \exp[(2 \pi i /L)(s+\phi(s))] =0\;.
\label{closed3}
\end{eqnarray}
However, since the model ignores the ${\cal O}(\phi^3)$ terms, this condition is not automatically fulfilled~\cite{Matsuo, Goldstein}. In order to overcome this problem, we employ the Bayesian estimate~\cite{Schervich}. First we replace the integrand of eq.  (\ref{closed3}) by $\exp[(2 \pi i /L)(s+\phi(s)+\epsilon(s))] $ where $\epsilon(s)$ is the unknown error function. We introduce the likelihood function as well as the {\it a priori} distribution for the modes of $\epsilon(s)$. By 
optimizing  the logarithmic likelihood, we determine $\epsilon(s)$ to enclose the domain boundary. 
 
For an internal consistency, we have verified numerically that the length $L$ is actually unchanged appreciably by these numerical methods. It is noted that the domain area is time-dependent in this model system.

\begin{figure}[t]
  \begin{center}
  \includegraphics[width=8.0cm]{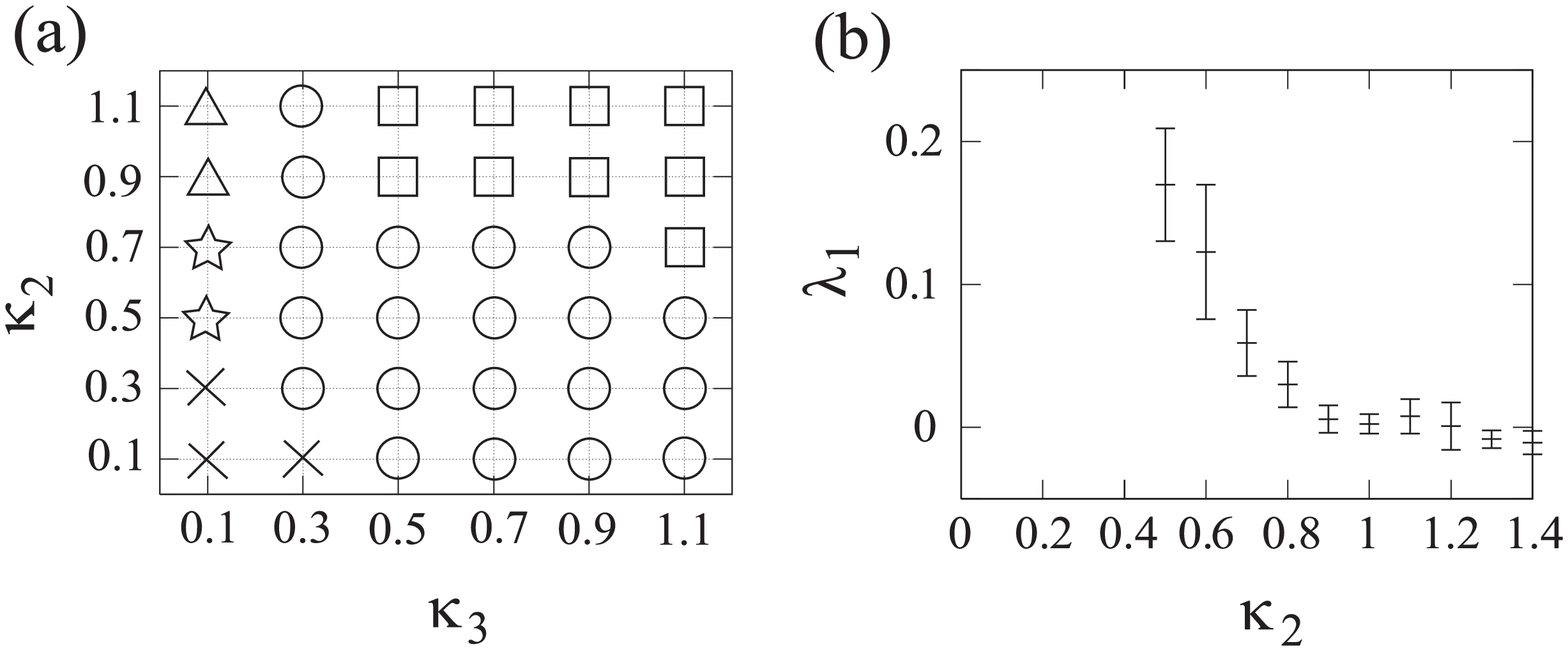} 
  \caption{(a) Phase diagram for the full set of equations
for $a_1=-1.0$, $b_1=-0.5$, $b_2=0.3$, $d_1=0.1$, $d_2=0.8$ and $\gamma=1.0$. 
The meaning of the symbols is given in the text.
In the region indicated by the symbol $\times$, an numerical instability occurs so that we have no definite conclusion about the motion. (b) Maximum Lyapunov exponent obtained numerically for $\kappa_3 = 0.1$. The other parameters are the same as those of the phase diagram (a). 
}
  \label{fig:phasediagram3}
  \end{center}
\end{figure}

As is seen in Figs. \ref{fig:phasediagram2} and  \ref{fig:phasediagram3}, the tensor model has exhibited several types of deformation dynamics by changing the three parameters $\gamma$, $\kappa_2$ and $\kappa_3$. In contrast, the active cell model produces these similar dynamics  changing only the value of $L$. Numerical simulations of Eqs. (\ref{phase_equation}) and (\ref{phase_velocity2}) 
show that the intricate dynamics appear in the three characteristic windows;
$W_1 \simeq \{L | L \in [12.0, 14.4]\}, W_2 \simeq \{ L | L \in [21.0, 21.9]\}$, and $W_3 \simeq \{ L | L \in [29.0, 29.9]\}$.
Circular, quasi-periodic, and rectangular motions emerge in $W_1$~\cite{Matsuo},
quasi-periodic and chaotic motions in $W_2$,
and zig-zag motions in $W_3$. These motions occur successively by changing the value of $L$.
The straight motion is obtained in the region $6.2 <L< 12.0$.  No shape instability occurs for $L<6.2$ and therefore no propagation of domain.
There is an obvious  reason why the complex dynamics appear in the windows $W_1$, $W_2$ and $ W_3$. 
The linear stability analysis of Eqs. (\ref{phase_equation}) about the trivial solution $\phi=0$ shows the Fourier mode-$n$ deformation becomes unstable at $L=2 \pi n$.
Actually the mode-2 becomes unstable in
$W_1=\{L| L \sim 4 \pi\}$, the mode-3 in $W_2=\{L| L \sim 6 \pi\}$ and the mode-4 in $W_3=\{L |L \sim 8 \pi\}$. Since the codimension two bifurcation points exist near these critical points,  the various types  of motion would appear.

\begin{figure}[t]
  \begin{center}
  \includegraphics[width=8.0cm]{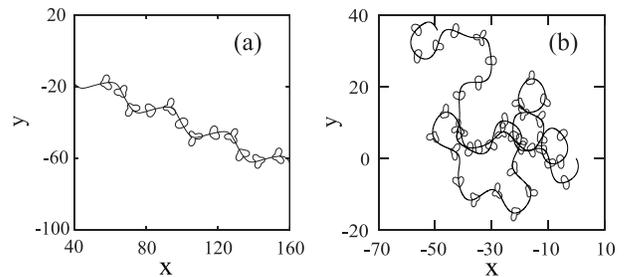} 
  \caption{(a) Zigzag motion for $\kappa_2=0.9$ and $\kappa_3=0.1$. The domain moves from the left to the right. (b) Chaotic motion for $\kappa_2=0.5$ and $\kappa_3=0.1$.}
  \label{fig:zigzag}
  \end{center}
\end{figure}

\begin{figure}[t]
\begin{center}
\includegraphics[scale=0.33]{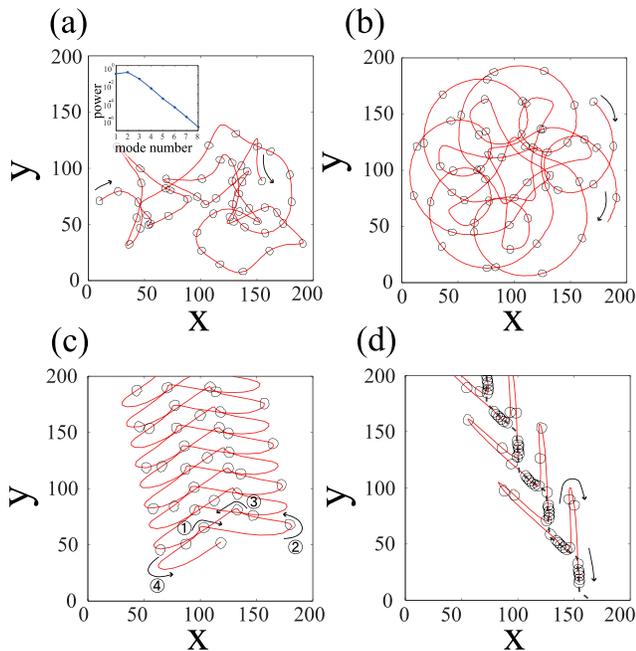}
\caption{Trajectories of a domain in the active cell model for $\alpha_3=-4\times 10^{-4}L$.
(a) Chaotic motion for $L=21.6$.
The inset shows the power spectrum calculated from the deformation modes.
(b) Quasi-periodic motion for $L=21.8$.
(c) Zig-zag motion for $L=29.2$ with the turn angle $180^\circ \pm 45^\circ$.
The circled numbers mean the qualifying order of trajectory. 
(d) Zig-zag motions for $L=29.6$.
The "coarse-grained" trajectory (the dotted gray line) shows $\pm 45^\circ$ turns.
}
\label{zigzag4}
\end{center}
\end{figure}

Figure \ref{zigzag4}(a) displays the trajectory obtained in  $W_2$,
where  an apparently  chaotic motion appears.
The broad power spectrum of the Fourier amplitudes of $\phi$ shown in the inset implies that the time-evolution of $\phi(s, t)$ is a kind of spatio-temporal chaos.
Figure \ref{zigzag4}(b) shows a quasi-periodic solution obtained in $W_2$.
Figures \ref{zigzag4}(c) and (d) show the two types of trajectory obtained in $W_3$.
Although the trajectory in Fig. \ref{zigzag4}(c) seems complicated, 
the orderly turns occurring in the trajectory ($\textcircled{\scriptsize1}, \textcircled{\scriptsize3} $, 
in Fig. \ref{zigzag4}(c)) have almost $180^\circ \pm 45^\circ$.
Thus this is a kind of zig-zag motion, however 
unlike the zig-zag motion in Fig. \ref{fig:zigzag}(a),
the effective modes governing the deformation are 
$n=\pm 4$ here.
The trajectory in Fig. \ref{zigzag4}(d) is also considered to be a kind of zig-zag motion,
because the main part of the trajectory indicated by the dotted gray line shows $\pm 45^\circ$ turns.
The zig-zag motion with $ n=\pm3$ as in Fig, \ref{fig:zigzag}(a) has not been found in the active cell model.

\section*{Coupled set of equations for the Fourier amplitudes}

The set of equations (\ref{eq:vtdyn}), (\ref{eq:Stdyn}) and (\ref{eq:Utdyn}) can be written in terms of the Fourier components in  eq. (\ref{eq:deltaR}).  
Here we define the complex variables  as
$z_1=v_1-iv_2$,
$z_2=\frac{1}{2}(S_{11}-iS_{12})$
and 
$z_3=\frac{1}{2}(U_{111}+iU_{222})$.
 The time-evolution equations for  $z_1$, $z_2$ and $z_3$ are given from eqs.  (\ref{eq:vtdyn}),  (\ref{eq:Stdyn})  and (\ref{eq:Utdyn}) by
\begin{eqnarray} \label{eq:dz1}
 \dot{z_1}&=& (\gamma + d_{11}|z_1|^2 + d_{12} |z_2|^2+ d_{13} |z_3|^2) z_1  \nonumber \\
 &+&  e_{11} \bar{z_1} z_2+e_{12}\bar{z_1}^2z_3 +e_{13}\bar{z_2}z_3
  + e_{14}\bar{z_3}z_2^2\;,
 \end{eqnarray}
\begin{eqnarray}  \label{eq:dz2}
 \dot{z_2}&=& (-\kappa_2 + d_{21}|z_1|^2 + d_{22} |z_2|^2+ d_{23} |z_3|^2) z_2 \nonumber \\
&+&  e_{21} z_1^2 +e_{22}\bar{z_1}z_3 
 + e_{23}\bar{z_2}z_3 z_1\;,
 \end{eqnarray}
\begin{eqnarray} 
\label{eq:dz3}
  \dot{z_3} &=& (-\kappa_3 + d_{31}|z_1|^2 + d_{32} |z_2|^2+ d_{33} |z_3|^2) z_3 \nonumber \\
&+&  e_{31} z_1^3 +e_{32}z_1z_2
 + e_{33}\bar{z_1}z_2^2 \;,
\end{eqnarray}
where the dot means the time derivative and the bar indicates the complex conjugate. All the coefficients are real and are given by
$d_{11}=-1$, $d_{12}=-8a_4$, $d_{13}=-16a_5$, 
$d_{21}=-b_4$, $d_{22}=-8b_3$, $d_{23}=-16b_5$,
$d_{31}=-d_4$, $d_{32}=-8d_5$, $d_{33}=-16d_3$,
$e_{11}=-2a_1$, $e_{12}=-2a_2$, $e_{13}=-8a_3$,
$e_{14}=16 a_6$, 
$e_{21}=b_1/4$, $e_{22}=b_2$, 
$e_{23}=2 b_6$, 
$e_{31}=d_1/8$, $e_{32}=d_2/2$
and $e_{33}=2 d_6$.
It is important to note that this set of equations is invariant under the transformation 
\begin{equation} \label{eq:trans1}
(z_1,z_2, z_3) \rightarrow (e^{i\theta}z_1 ,e^{2i\theta}z_2, e^{3i\theta}z_3)\;,
\end{equation}
for an arbitrary phase angle $\theta$. 
This invariance arises from the isotropy of space.

When the variable $z_3$ is omitted, 
the set of equations (\ref{eq:dz1}) and (\ref{eq:dz2}) is the same as those considered by Armbruster et al \cite{Arm1}. 
They  motivated to study some partial differential equation like the Kuramoto-Sivashinsky equation in one dimension under 
the periodic boundary condition \cite{Arm2} and had no consideration of the self-propelled domain dynamics.  
In fact, there are the following correspondences; the motionless circular shape domain $\leftrightarrow$ the trivial solution, 
the deformed motionless domain $\leftrightarrow$ pure mode, the straight motion $\leftrightarrow$ the standing wave, 
the rectangular motion $\leftrightarrow$ the heteroclinic cycle, the rotating motion $\leftrightarrow$ the traveling wave 
and the quasi-periodic motion $\leftrightarrow$ the modulated wave. The former is the motions 
obtained in our theory whereas the latter is the terminology of Armbruster et al \cite{Arm1}.

The variable $\phi$ in the active cell model can also be represented in terms of the Fourier modes.  If the modes higher than $n=3$ relax rapidly to the stationary values, we may retain only the modes  $n=1, 2, 3$. The time-evolution equations are essentially the same structures as those in eqs. (\ref{eq:dz1}), (\ref{eq:dz2}) and (\ref{eq:dz3}). Consequences of this will be discussed below.

\section*{Discussion}
We have shown that various self-propelled motions appear both in the tensor model and the active cell model. The dynamics common to these two systems are straight motion, circular periodic and quasi-periodic motions, rectangular motion, zig-zag motions and chaotic motion.

Here we note the main difference between our model and the
self-propelled swimmers at low Reynolds number\cite{Percell}. Our model
equations are autonomous, thus the shape deformation and the centroid
migration are spontaneously created.  On the other hand in the latter frameworks, the deformation of flexible body is operationally given and resulting motion is considered within Stokes dynamics. 
Owing to the autonomous properties, our models exhibit successive bifurcations leading to richer dynamics. Since the tensor model 
 (\ref{eq:vtdyn}), (\ref{eq:Stdyn}), and  (\ref{eq:Utdyn}) have been derived from the excitable reaction diffusion model,  the bifurcation from a simple to a complex motion should be explored experimentally in physico-chemical systems such as oily droplet systems. 
 Along this line, propagating actin waves and recovery of actin polymerization from complete depolymerization
observed in Dictyostelium cells\cite{Gerisch} might give a clue to a connection between our model and possible bifurcations in the dynamics of living cells.

Moreover, we note the difference of the zigzag motions in the two models. The zig-zag motion with the angle about $60^\circ$ has been obtained in the tensor model. This is attributed to the fact that only the second and the third modes are considered in the tensor model so that the deformation with three fold symmetry is possible which triggers the $60^\circ$ zigzag motion.
In the active cell model, on the other hand, more complicated zigzag motion with the angle about $135^\circ$ and $45^\circ$ appears as 
in Figs.  \ref{zigzag4}(c) and (d) where $L/2\pi$ is close to 4. Therefore, it is expected that higher modes such as the fourth mode are dominant for the zig-zag motion in  the active cell model.

We emphasize that the rectangular motion, the $60^\circ$ zigzag motion and an apparently chaotic motion have been observed in real experiments of amoebas \cite{Cox, Maeda2008}. Therefore the present approach based on the symmetry argument to construct the time-evolution equations captures the essential feature of the coupling between the shape and the motion of a self-propelled domain.

\section*{Acknowledgment}
This work was supported by 
the Grant-in-Aid for priority area "Soft Matter Physics" 
from the Ministry of Education, Culture, Sports, Science and Technology (MEXT) of Japan.


\begin{thebibliography}{99}
\bibitem{Taylor}
TAYLOR G., {\it Proc. R. Soc. Lond. A}, {\bf 211} (1952) 225. 
\bibitem{Percell}
PERCELL E. M., {\it Am. J. Phys.}, {\bf 45} (1977) 3. 
\bibitem{Ishikawa}
ISHIKAWA T. and PEDLEY, T. J., {\it J. Fluid Mech.}, {\bf 588} (2007) 399. 
\bibitem{Ram}
HATWALNE Y. {\it et al.}, {\it Phys. Rev. Lett.},  {\bf 92} (2004) 118101.
\bibitem{Wada}
WADA H. and NETZ R.R.,  {\it Phys. Rev. Lett.}, {\bf 99}  (2007) 108102.
\bibitem{Yeomans}
ALEXANDER G. P. and YEOMANS J. M., {\it Europhys. Lett.}, {\bf 83} (2008) 34006.
\bibitem{Sumino}
SUMINO Y. {\it et al.}, {\it Phys. Rev. Lett.},  {\bf 94} (2005) 068301.
\bibitem{Nagai}
NAGAI K. {\it et al.},  {\it Phys. Rev. E}, {\bf 71}  (2005) 065301(R).
\bibitem{Sugawara2009}
SUZUKI K. {\it et al.}, {\it Chemistry Letters}, {\bf 38} (2009) 1010. 
\bibitem{Ajdari}
GOLESTANIAN R. {\it et al.},  {\it New J. Phys.}, {\bf 9}  (2007) 126.
\bibitem{Nishimura}
NISHIMURA S. I. {\it et al.}, {\it PLoS Comput. Biol}, {\bf 5} (2009) e1000310.
\bibitem{Kapral}
TAO Y.-G. and KAPRAL R.,  {\it J. Chem. Phys.}, {\bf 131} (2009) 024113/
\bibitem{Cox}
LI L., NORRELYKKE S. F. and COX E. C., {\it PLoS one}, {\bf 3} (2008) e2093.
\bibitem{Maeda2008}
MAEDA Y. T. {\it et al.}, {\it PLoS one}, {\bf 3} (2008) e3734. 
\bibitem{OhtaOhkuma}
OHTA T. and OHKUMA T., {\it Phys. Rev. Lett.}, {\bf 102} (2009) 154101.
\bibitem{Mikhailov}
KRISCHER K. and MIKHAILOV A., {\it Phys. Rev. Lett.}, {\bf 73} (1994)  3165.
\bibitem{OOS}
OHTA T., OHKUMA T. and SHITARA K., {\it Phys. Rev. E}, {\bf 80} (2009) 056203. 
\bibitem{Matsuo}
MATSUO M. Y., MAEDA Y. T. and SANO M., unpublished.
\bibitem{Goldstein}
GOLDSTEIN R. and LANGER S. A., {\it Phys. Rev. Lett.}, {\bf 75} (1995) 1094.
\bibitem{Arm1}
ARMBRUSTER D., GUCKENHEIMER J. and HOLMES P., {\it Physica D}, {\bf 29} (1988) 257. 
\bibitem{Arm2}
ARMBRUSTER D., GUCKENHEIMER J. and HOLMES P., {\it SIAM J. Appl. Math.}, {\bf 49} (1989) 676.
\bibitem{Schervich}
SCHERVISH M. J., {\it Theory of Statistics} (Springer-Verlag, New York) 1995.
\bibitem{Gerisch}
GERISCH G. {\it et al.}, {\it Bio. Phys. J.}, {\bf 87} (2004) 3493.

\end{thebibliography}
\end{document}